\documentstyle[11pt,newpasp,twoside]{article}

\begin{document}

\title{Gas in the central region of AGNs: ISM and supermassive gaseous objects}
\author{P. Amaro-Seoane and R. Spurzem}
\affil{Astronomisches Rechen-Institut, M\"onchhofstr. 12-14\\
Heidelberg D-69120, Germany\\
email: pau, spurzem@ari.uni-heidelberg.de}

\begin{abstract}
This is a starting point of future work on a 
more detailed study of early evolutionary phases 
of galactic nuclei. We put into port the study of 
the evolution of star accretion onto a supermassive 
gaseous object in the central region of an active galactic
nucleus, which was previously addressed semi-analytically. For
this purpose, we use a gaseous model of relaxing dense
stellar systems, whose equations are solved numerically.
The model is shortly described and some first model calculations
for the specific case of black hole mass growth in time are given.
Similar concepts will be applied in future work for a system
in which there is a dense interstellar medium as a possible
progenitor of a black hole in the galactic centre.

\end{abstract}

\section{Introduction}
Several theoretical models have been proposed in order to explain
the properties of quasars and other types of active galactic nuclei (AGN). 
In the 60's and
70's supermassive central objects (SMOs) were thought to be the main source 
of their characteristics (luminosities of $L \approx 10^{12} {\rm L}_{\odot
}$ produced on very small scales, jets etc). The release of gravitational
binding energy by the accretion of material onto an SMO in the range of
$10^7 - 10^9 {\rm M}_{\odot}$ has been suggested to be the primary powerhouse
(Lynden-Bell, 1969). Supermassive stars (SMSs) and supermassive black holes
(SMBHs) are two possibilities to explain the nature of these SMOs, and the
first may be an intermediate step towards the formation of the former type 
(Rees 1984).
Large amounts of gas lost by stars during their evolution will stock in 
galactic centres, as have shown numerical studies of the evolution of gas
in the bulge of spiral galaxies (Loose et al. 1982) and in elliptical
(Loose and Fricke 1980, Kunze et al. 1987). 

In previous work we addressed a semi-analytical study, revisiting and
expanding classical paper's work, the classical problem of star 
accretion onto a supermassive central gaseous object in a galactic nucleus
(Amaro-Seoane \& Spurzem 2001). The resulting supermassive central gas-star 
object was assumed to
be located at the centre of a dense stellar system for which we used a 
simplified model consisting of a Plummer model with an embedded density 
cusp using stellar point masses. From the number of stars belonging to 
the loss-cone, which plunge onto the central object on elongated orbits 
from outside, we estimated the accretion rate taking into account a possible 
anisotropy of the surrounding stellar distribution. 

This initial approximation to the real physical configuration
was just a first probe of the complexity of the problem. 
We envisaged the problem from 
a static point of view. This allowed us to treat it
in a semi-analytical way. Such a treatment is (of course) not 
realistic, but just a useful approach, for it gives us a nice and
simple basic idea of what could be going on in such a particular 
scenario. In this paper we 
put into port the evolution of the static situation thanks to an anisotropic 
gaseous model.
A next step will include the implementation of the small-scale 
fluctuations on the large-scale motion of the gas in order to study the 
transfer of momentum between the interstellar medium (ISM) and the system
of stars. Just et al. (1986) derived an equation for the large-scale motion
of the interstellar medium in the presence of small-scale fluctuations.
Of particular interest is the study of the dynamical friction between 
the ISM and the system of stars, which 
includes the development of a friction term describing
the momentum transfer between the interstellar gas and the 
stellar component.

\section{The gaseous model}

To go from stationary to dynamical models we use a gaseous model of
star clusters (Lynden-Bell \& Eggleton 1980, Bettwieser 1983, Heggie 1984)
in its anisotropic version (Louis \& Spurzem 1991, Spurzem 1994, Giersz \& Spurzem 1994).
It is based on the following basic assumptions:
\begin{itemize}

\item The system can be described by a one particle distribution function.

\item The secular evolution is dominated by the cumulative 
effect of small angle deflections with small impact parameters (Fokker-Planck
approximation, good for
large $N$-particle systems). 

\item The effect of the two-body relaxation can be modelled by a local
heat flux equation with an appropriately tailored conductivity.

\end{itemize}
\noindent
The first assumption justifies a kinetic equation of the Boltzmann type
with the inclusion of a collisional term of the Fokker-Planck (FP) type:

\begin{equation}
\frac{\partial f}{\partial t}+v_{r}\frac{\partial f}{\partial r}+\dot{v_{r}}
\frac{\partial f}{\partial v_{r}}+\dot{v_{\theta}}\frac{\partial f}{\partial 
v_{\theta}}+\dot{v_{\varphi}}\frac{\partial f}{\partial v_{\varphi}}=\Bigg
( \frac{\delta f}{\delta t} \Bigg)_{FP}
\end{equation}
\noindent
In spherical symmetry polar coordinates $r$, $\theta$, $\phi$
are used and $t$ denotes the time. The vector
${\bf v} = (v_i), i=r,\theta,\phi$ denotes the velocity
in a local Cartesian coordinate system at the spatial point
$r,\theta,\phi$. The distribution function $f$ is a function of,
$r$, $t$, $v_r$, $v_\theta^2+v_\phi^2$ only due to spherical symmetry.
By multiplication of the Fokker-Planck equation (1) with
various powers of the velocity components we get up to second order a
set of moment equations which is equivalent to gas-dynamical equations
coupled with Poisson's equation: A mass equation, a continuity equation,
an Euler equation (force), radial and tangential energy equations.
The system of equations is closed by a phenomenological heat flux equation
for the flux of radial and tangential r.m.s. kinetic energy, both in
radial direction. The concept is physically very similar to the original
one of Lynden-Bell \& Eggleton (1980). As a consequence we derive a
set of dynamical evolution equations in spherical symmetry for our system
which resemble very much standard gas dynamical equations. The reader
interested in more details is referred to the above cited papers.
A simple diffusion model is then used to describe the diffusion
of stars into the loss cone and their subsequent accretion onto
the black hole, which will be published soon (Amaro-Seoane \& Spurzem,
in preparation).

\begin{figure}[t!]
\plotfiddle{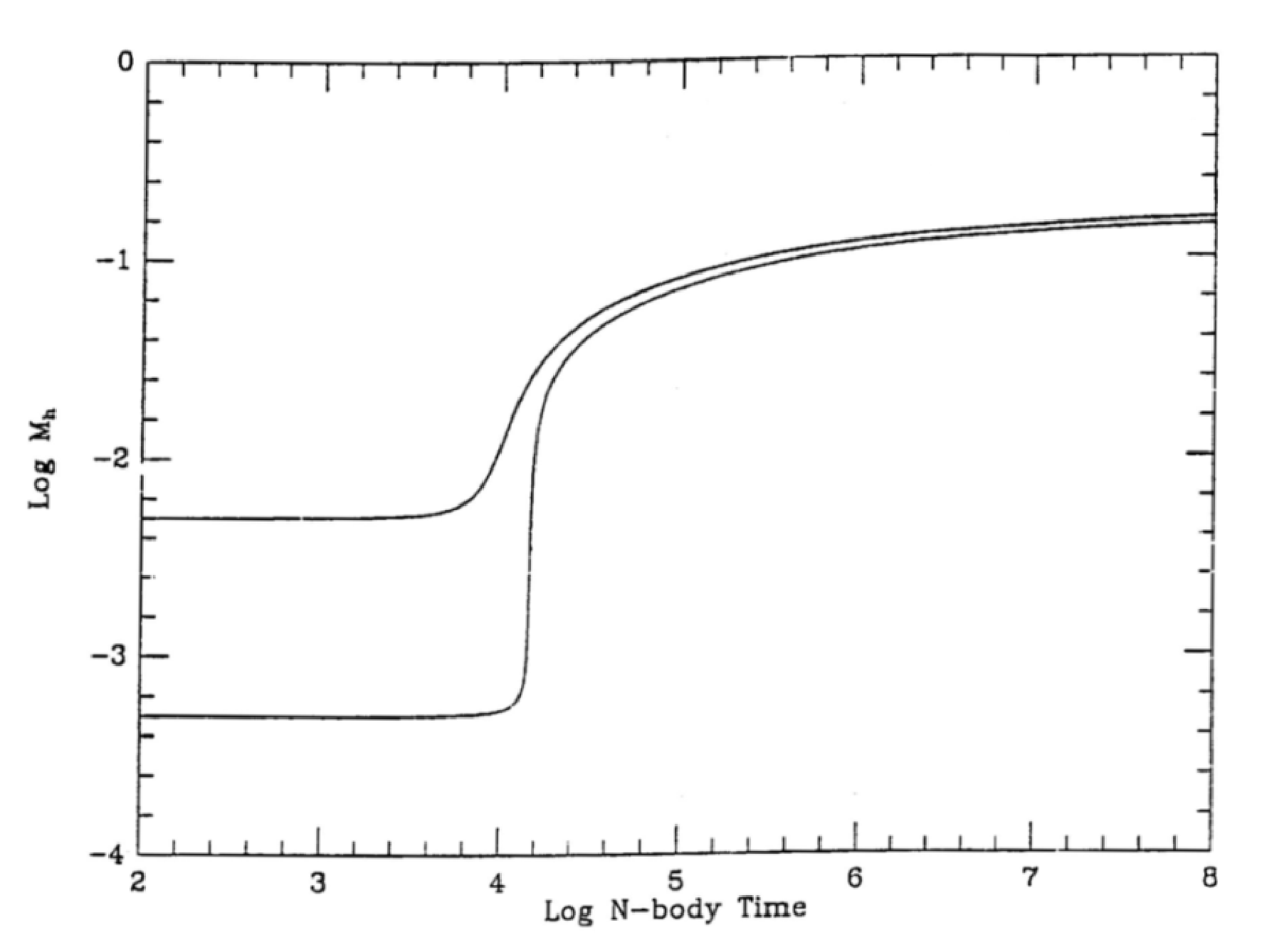}{8.0cm}{0}{30}{30}{-140}{-20}
\caption{Central Black Hole mass in ${\rm M}_\odot$ as a function of time
(1 time unit corresponds to $1.04\cdot 10^{12}$ sec. The lower curve starts
at a seed mass of 50, the upper at 500 ${\rm M}_\odot$, respectively.}
 \end{figure}

To solve our equations we discretise them on a logarithmic radial mesh 
with typically 200 mesh points, covering radial scales over eight orders
of magnitude (e.g. from 100 pc down to $10^{-6}$ pc, which is enough to
resolve the system down to the vicinity of a massive black hole's tidal
disruption radius for stars. An implicit
Newton-Raphson-Henyey iterative method is used to solve for the time
evolution of our system.
\vskip0.3cm
As a sample, in Fig. 1 we follow the evolution 
of the mass of a black hole (BH) in an isolated star cluster of $10^5 {\rm M}_\odot$ 
undergoing core collapse. We start with
two initial masses for the BH's of 50 and 500 (normalised to the total cluster mass), the 
time unit in the figure
is $1.04\cdot 10^{12}$ sec. The central density increases until a maximum value
at which the energy provided by the star accretion onto the black hole
stops the collapse; the subsequent reexpansion of the cluster core
halts the black hole growth. Independent of the initial seed black
hole mass the final mass is very similar, which we consider as an
effect of self-regulation due to the feedback of black hole growth
to the cluster evolution. 

Our results compare well with other studies
using direct solutions of the Fokker-Planck equation or Monte Carlo models
(Lightman \& Shapiro 1977, Marchant \& Shapiro 1980). Note that the
Monte Carlo approach has been recently revisited and improved by
Freitag \& Benz (2001). In contrast to the other models the gaseous
model is much more versatile to include all kinds of important other
physical effects, such as the dynamics of gas liberated in nuclei
by stellar evolution and collisions and its interaction with the
stellar component (see for a preliminary study Langbein et al. 1990).

\end{document}